\documentclass[twocolumn,showpacs,preprintnumbers,amsmath,amssymb,aps]{revtex4-1}
\usepackage{graphicx}% Include figure files
\usepackage{dcolumn}% Align table columns on decimal point
\usepackage{bm}% bold math

\begin{document}

\title{Ytterbium-driven strong enhancement of electron-phonon coupling in graphene}
\author{Choongyu Hwang$^{1,2}$}
\author{Duck Young Kim$^{3}$}
\author{D. A. Siegel$^{1,4}$}
\author{Kevin T. Chan$^{1,4}$}
\author{J. Noffsinger$^{1,4}$}
\author{A. V. Fedorov$^{5}$}
\author{Marvin L. Cohen$^{1,4}$}
\author{B\"{o}rje Johansson$^{6}$}
\author{J. B. Neaton$^{4,7,8}$}
\author{A. Lanzara$^{1,4}$} \email{ALanzara@lbl.gov}
\affiliation{$^1$Materials Sciences Division, Lawrence Berkeley
National Laboratory, Berkeley, CA 94720,
USA}\affiliation{$^2$Department of Physics, Pusan National
University, Busan 609-735, Republic of
Korea}\affiliation{$^3$Geophysical Laboratory, Carnegie Institution
of Washington, Washington DC 20015, USA} \affiliation{$^4$Department
of Physics, University of California, Berkeley, CA 94720, USA}
\affiliation{$^5$Advanced Light Source, Lawrence Berkeley National
Laboratory, Berkeley, CA 94720, USA} \affiliation{$^6$Department of
Materials and Engineering, Royal Institute of Technology, Stockholm,
SE-100 44, Sweden} \affiliation{$^7$The Molecular Foundry, Lawrence
Berkeley National Laboratory, Berkeley, CA 94720, USA}
\affiliation{$^8$Kavli Energy Nanosciences Institute at Berkeley,
Berkeley, CA 94720, USA.}

\date{\today}

\begin{abstract}
We present high-resolution angle-resolved photoemission spectroscopy
study in conjunction with first principles calculations to
investigate how the interaction of electrons with phonons in
graphene is modified by the presence of Yb. We find that the
transferred charges from Yb to the graphene layer hybridize with the
graphene $\pi$ bands, leading to a strong enhancement of the
electron-phonon interaction. Specifically, the electron-phonon
coupling constant is increased by as much as a factor of 10 upon the
introduction of Yb with respect to as grown graphene ($\leq$0.05).
The observed coupling constant constitutes the highest value ever
measured for graphene and suggests that the hybridization between
graphene and the adatoms might be a critical parameter in realizing
superconducting graphene.
\end{abstract}

\pacs{71.38.-k,72.10.Di,73.20.-r,79.60.-i}
%71.38.-k Polarons and electron-phonon interactions
%72.10.Di Scattering by phonons, magnons, and other nonlocalized excitations
%73.20.-r Electron states at surfaces and interfaces
%79.60.-i Photoemission and photoelectron spectra

\maketitle

\section{Introduction}
The interaction of electrons with phonons is of practical and
fundamental interest in graphene, as it not only affects the
transport properties of actual devices~\cite{Efetov}, but also
induces novel phenomena such as charge density waves~\cite{Rahnejat}
and superconductivity~\cite{Uchoa}. Hence the manipulation of the
electron-phonon coupling is an important issue to realize
graphene-based electronic and spintronic devices~\cite{Geim} and to
create new strongly correlated electron phases. In fact, several
methods have been proposed to modify the electron-phonon coupling
constant, $\lambda$, of graphene using charge carrier
density~\cite{Calandra_resolution}, magnetic field~\cite{Faugeras},
disorder~\cite{Song}, and adatoms~\cite{Profeta}. Among them, the
change of charge carrier density can tune the strength of
electron-phonon coupling up to
$\lambda$$\leq$0.05~\cite{Calandra_resolution}, while
electron-electron interactions are efficiently
suppressed~\cite{DavidNJP}. On the other hand, the presence of
adatoms is predicted to drastically enhance electron-phonon coupling
up to $\lambda$=0.61~\cite{Profeta}, so that graphene enters the
regime where phonon-mediated superconductivity might
exist~\cite{Savini,Profeta}. However, experimental evidence of this
striking enhancement in graphene has been controversial so far.

The most prominent manifestation of the electron-phonon coupling is
a renormalization or kink of the electronic band structure at the
phonon energy accompanied by a change in the charge carrier
scattering rate. These effects are directly observed using
angle-resolved photoemission spectroscopy (ARPES)~\cite{AM}.
However, experimental studies on the role of adatoms for the
electron-phonon coupling of graphene via ARPES have been debated due
to the hybridization of the adatom band with the graphene $\pi$
bands, referred to as band structure effect, resulting in an
apparent enhancement and anisotropy of the electron-phonon coupling
strength~\cite{Jessica,Calandra_BandStructureEffect,Park_vanHove}.
On the other hand, strong enhancement of the electron-phonon
coupling through adatom intercalation have been reported for
graphite, and discussed in same cases as the driver for
superconductivity~\cite{Valla2009,Yang}. These previous results
suggest the importance of combining experimental and theoretical
studies to understand the enhancement of electron-phonon coupling in
graphene.

Here we present high-resolution ARPES study showing a strong
enhancement of the electron-phonon coupling strength in a monolayer
graphene sheet via Yb adsorption. A direct comparison with the
theoretical band structure determined by first principles
calculations show that the Yb 6$s$ electrons transferred to the
graphene layer are hybridized with the graphene $\pi$ bands,
resulting in an enhanced electron-phonon coupling from
$\lambda$=0.05 for as grown graphene to $\lambda$=0.43 for graphene
with Yb. This observation constitutes the highest value ever
measured for graphene and is in line with the density-functional
perturbation theory that predicts an enhancement of $\lambda$ from
0.02 to 0.51.

\section{Methods}
\subsection{Experimental details}
%-------------------Figure 1: intro ------------------------%
\begin{figure*}
\begin{center}
\includegraphics[width=1.9\columnwidth]{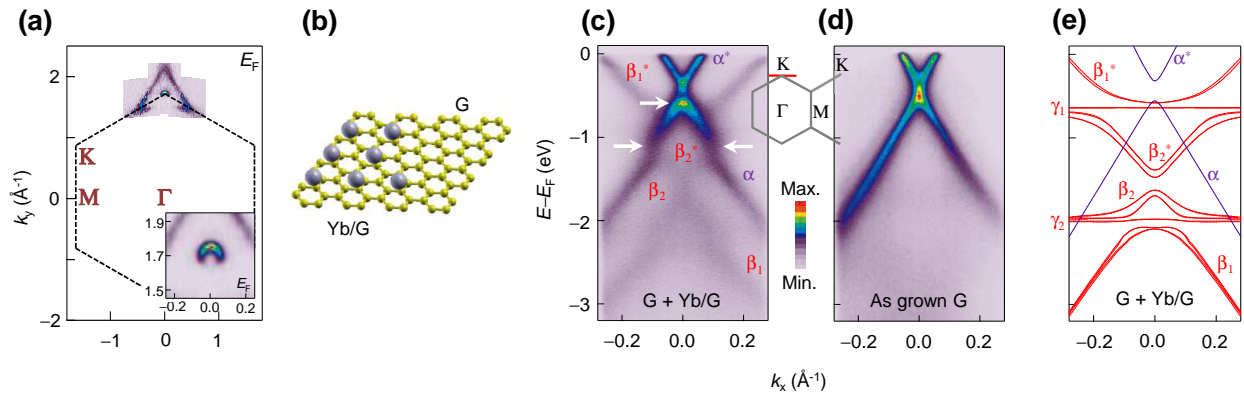}
\end{center}
\caption{\label{fig:fig1} %{\bf Electronic band structure of Yb-adsorbed graphene}
(Color online) (a) Fermi surface of graphene upon the introduction
of Yb. Inset shows a zoomed-in view near the K point ($k_{\rm x}$,
$k_{\rm y}$)=(0, 1.7). (b) Schematic drawing of crystal structure of
graphene with inhomogeneous contribution of Yb (the buffer layer and
the SiC substrate are not drown for simplicity). (c-d) Raw ARPES
data of graphene in the presence of Yb (G+Yb/G: panel (c)) and as
grown graphene (As grown G: panel (d)), through the K point
perpendicular to the $\Gamma$K direction as denoted by the red line
in the inset. (e) Calculated bands of graphene in the presence of Yb
(G$+$Yb/G) perpendicular to the $\Gamma$K direction. The G and Yb/G
bands are purple and red curves, respectively. The graphene $\pi$
bands are denoted by $\alpha$ and $\alpha^{\ast}$, and the Yb/G
$\pi$ bands by $\beta$ and $\beta^{\ast}$. Yb 4$f_{7/2}$ and
4$f_{5/2}$ electrons are denoted by $\gamma_1$ and $\gamma_2$. }
\end{figure*}
%-----------------------------------------------------------------

Single layer graphene was grown epitaxially on a 6{\it H}-SiC(0001)
substrate by an e-beam heating method as described
elsewhere~\cite{Rolling}. Yb was deposited on graphene at 100~K,
followed by repeated annealing processes from 400~K to 1000~K to
find a stable geometric structure.  This process is well-known to
enhance intercalation of alkali- and alkali-earth metals such as K
and Ca~\cite{Jessica}, Rb and Cs~\cite{Watcharinyanon2011}. This is
also true for Yb~\cite{Watcharinyanon2013_a,Watcharinyanon2013_b}
when annealed above $\sim$200~$^{\circ}$C. As a result, the graphene
sample in the presence of Yb exhibits a coexisting phase of
Yb-intercalated graphene and graphene without Yb, as observed in
Fig.~\ref{fig:fig1}(a). High-resolution ARPES experiments were
performed at beamline 12.0.1 of the Advanced Light Source in
ultra-high vacuum maintained below 2$\times$10$^{-11}$~Torr using a
photon energy of 50~eV. The energy and angular resolutions were
32~meV and $\leq$0.2~$^{\circ}$, respectively. The measurement
temperature was 15~K.

\subsection{Electronic band structure calculations}
%The Yb/G bands are obtained for YbC$_6$ by {\it ab initio}
%pseudopotential total energy calculations with plane-wave basis
%set~\cite{Cohen}. The exchange-correlation of electrons was treated
%within the generalized gradient approximation (GGA) as implemented
%by Perdew-Berke-Enzelhof~\cite{Perdew}. The on-site Coulomb ($U$)
%and intra-atomic exchange ($J$) interactions are determined to be
%2.0~eV and 0.7~eV, by varying $U$ until a good agreement between
%measured and calculated bands is obtained using the $GGA+U$
%correction to the $f$ electrons of Yb. These values are different
%from 5.4~eV and 0.7~eV for Yb-intercalated graphite extracted from
%the full potential linear augmented plane wave method (LAPW) with
%$LDA+U$ correction~\cite{Mazinb}. The $U$ value calculated within
%$LDA+U$ scheme is usually overestimated due to the confined
%screening charge in the same atomic sphere~\cite{Mazinb}. Although
%it is not straightforward to directly compare $U$ values estimated
%by two different correction methods, Yb/G shows smaller value than
%that of Yb-intercalated graphite.
The electronic band structure of graphene with Yb are obtained for
YbC$_6$ by \textit{ab initio} total energy calculations with a
plane-wave basis set~\cite{Cohen} performed using the Vienna
Ab-initio Simulation Package (VASP) \cite{PhysRevB.47.558,
Kresse1996, PhysRevB.54.11169}. Projector augmented wave (PAW)
potentials \cite{Blochl1994, Kresse1999} with a plane-wave cutoff of
500~eV are used. The exchange-correlation of electrons was treated
within the generalized gradient approximation (GGA) as implemented
by Perdew, Burke, and Ernzerhof~\cite{Perdew}. The comparison
between the measured and the calculated bands using GGA$+$U
correction to the $f$ electrons of Yb bears 2.0~eV for the on-site
Coulomb interaction (U) and 0.7~eV for the intra-atomic exchange
interaction (J)~\cite{Liechtenstein1995}. These values differ from
5.4~eV and 0.7~eV, respectively, expected for Yb-intercalated
graphite as extracted from the full potential linear augmented plane
wave method (LAPW) with $LDA+U$ correction~\cite{Mazinb}. The $U$
value calculated within $LDA+U$ scheme is usually an overestimate
due to the confined screening charge in the same atomic
sphere~\cite{Mazinb}. Although it is not straightforward to directly
compare $U$ values estimated by two different correction methods,
Yb/G shows smaller value than that of Yb-intercalated graphite.

\section{Results}
%-------------------Figure 2: hybridization ------------------------%
\begin{figure*}
\begin{center}
\includegraphics[width=1.9\columnwidth]{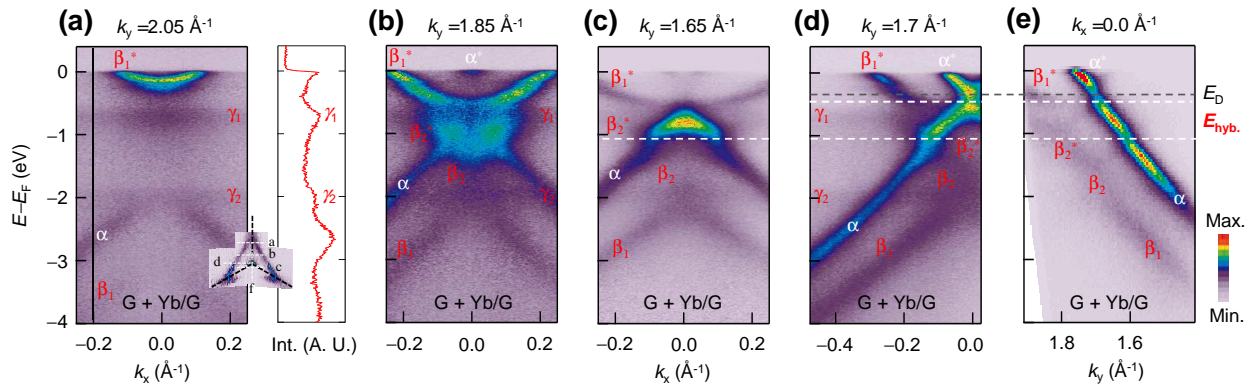}
\end{center}
\caption{\label{fig:fig2} (Color online) (a) Energy spectra of the
Yb/G band taken perpendicular to the $\Gamma$K direction at $k_{\rm
y}$=2.05~\AA$^{-1}$ denoted in the inset. The intensity spectrum is
taken at $k_{\rm x}$=$-$0.2~\AA$^{-1}$ denoted by the black solid
line. (b-d) Energy spectra of the Yb/G band taken perpendicular to
the $\Gamma$K direction at $k_{\rm y}$=1.85, 1.65, and
1.7~\AA$^{-1}$, respectively, denoted in the inset of panel (a). (e)
Energy spectra of the Yb/G bands parallel to the $\Gamma$K direction
denoted in the inset of panel (a). $E_{\rm hyb.}$ represents the
hybridization energy between the G ($\alpha$) and Yb/G
($\beta_2^{\ast}$ and $\beta_1^{\ast}$) bands, denoted by white
dashed lines, and $E_{\rm D}$ is the Dirac energy.}
\end{figure*}
%-----------------------------------------------------------------

Figure~\ref{fig:fig1}(a) shows a photoelectron intensity map at
$E_{\rm F}$ as a function of two dimensional wave vectors $k_{\rm
x}$ and $k_{\rm y}$, for graphene with Yb. Two pieces of Fermi
surface can be clearly distinguished: one with a crescent-like shape
centered at the Brillouin zone corner K (zoomed-in in the inset),
which resembles the one measured for as grown graphene on
SiC(0001)~\cite{Hwang2011}, and the other with a triangular shape
with the apex near the M point, similar to that of highly
electron-doped graphene~\cite{Jessica}. The observation of these two
Fermi surfaces suggests a coexistence of graphene with and without
Yb, as schematically shown in Fig.~\ref{fig:fig1}(b), similar to the
case of Rb- and Cs-adsorbed graphene~\cite{Watcharinyanon2011} and
consistent with previous results on Yb-intercalated
graphene~\cite{Watcharinyanon2013_a,Watcharinyanon2013_b}. An
estimate of the charge doping in the graphene $\pi$ bands introduced
by Yb is given by the area enclosed by the Fermi surface. The
occupied area for the crescent-like Fermi surface is
0.025~\AA$^{-2}$, which corresponds to an electron doping of
$n$$\sim$1.2$\times$10$^{13}$~cm$^{-2}$, similar to the one reported
for as grown graphene~\cite{DavidAPL}. The larger triangular Fermi
surface, which corresponds to an area of 0.33~\AA$^{-2}$, yields a
much higher electron doping of
$n$$\sim$1.7$\times$10$^{14}$~cm$^{-2}$. The electronic band
structure of the former crosses $E_{\rm F}$ at $k_{\rm
x}$=$\pm$0.063~\AA$^{-1}$ (spectra with the strongest intensity in
Fig.~\ref{fig:fig1}(c)) with a Dirac point at $\sim$$-$0.4~eV, which
resembles as grown graphene shown in Fig.~\ref{fig:fig1}(d), except
for the observed discontinuities around 0.6~eV and 1.1~eV below
$E_{\rm F}$ as denoted by white arrows in Fig.~\ref{fig:fig1}(c). On
the other hand, the electronic band structure of the latter crosses
$E_{\rm F}$ at $k_{\rm x}$=$\pm$0.26~\AA$^{-1}$ (spectra with the
weakest intensity in Fig.~\ref{fig:fig1}(c)) with a Dirac point at
$\sim$$-$1.6~eV.

Figure~\ref{fig:fig1}(e) shows the calculated electronic band
structure for the inhomogeneous sample, where closed packed islands
of YbC$_6$ (referred to as ``Yb/G'' bands) coexist with islands of
clean graphene without Yb (referred to as ``G'' bands). The G bands,
shown in purple and denoted by $\alpha$ and $\alpha^{\ast}$, are the
well known graphene $\pi$ bands obtained within the tight-binding
formalism~\cite{GruneisTB} in the presence of an energy gap of
0.2~eV at $E_{\rm D}$~\cite{Zhou2007,Kim2008}, while the origin of
the gap-like feature is still
controversial~\cite{Kim2008,Eli2008_comment,Zhou2008_comment,Eli_plasmaron,Louie}.
The Yb/G bands, shown in red and obtained by {\it ab initio}
pseudopotential total energy calculations with a plane-wave basis
set~\cite{Cohen}, are denoted by $\beta_1$, $\beta_1^{\ast}$,
$\beta_2$, $\beta_2^{\ast}$, $\gamma_1$, and $\gamma_2$. $\beta$ and
$\beta^{\ast}$ are the $\pi$ bands of the Yb/G, while $\gamma_1$ and
$\gamma_2$ are the Yb 4$f_{7/2}$ and 4$f_{5/2}$ electrons,
respectively. The Yb 4$f$ electrons are strongly hybridized with
$\beta^{\ast}$ and $\beta$ bands at 0.7~eV and 2.0~eV below $E_{\rm
F}$, respectively, resulting in a departure of the Yb/G band from
$\beta^{\ast}$ to $\beta_1^{\ast}$ and $\beta_2^{\ast}$, and from
$\beta$ to $\beta_1$ and $\beta_2$. The observed discontinuities at
the crossing points of $\alpha$ with $\beta_1^{\ast}$ and
$\beta_2^{\ast}$ (white arrows in Fig.~\ref{fig:fig1}(c)) may
indicate that the G and Yb/G are electronically coupled with each
other.

The $\gamma_1$ and $\gamma_2$ bands show weak spectral intensity
with respect to the other bands near the K point. Their relative
intensity is enhanced away from the K point, as shown in
Figs.~\ref{fig:fig2}(a-d), in which ARPES data were taken
perpendicular to the $\Gamma$K direction at several $k_{\rm y}$
values denoted in the inset of Fig.~\ref{fig:fig2}(a). The position
of $\gamma_1$ and $\gamma_2$ is determined by the intensity spectrum
at $k_{\rm x}$=$-$0.2~\AA$^{-1}$ denoted by a black solid line in
Fig.~\ref{fig:fig2}(a). The hybridization between the Yb/G and Yb
bands is clear at $k_{\rm y}$=1.85~\AA$^{-1}$ as shown in
Fig.~\ref{fig:fig2}(b). The deformation of the Yb/G band from
$\beta^{\ast}$ to $\beta_1^{\ast}$ and $\beta_2^{\ast}$ is observed
at the crossing points with the $\gamma_1$ band. The $\beta$ band
also shows unusual discontinuity at the crossing points with the
$\gamma_2$ band as shown in Figs.~\ref{fig:fig2}(b)
and~\ref{fig:fig2}(c). Such a hybridization is not observed between
the G band and Yb 4$f$ electrons, e.\,g.\,, $\alpha$ does not show
such a deformation or discontinuity at the crossing point with the
$\gamma_2$ band around ($E-E_{\rm F}$, $k_{\rm x}$)=($-$2.0,
$-$0.26) in Fig.~\ref{fig:fig2}(d). On the other hand, the
hybridization between the Yb/G and G bands is clear from the energy
spectra not only along $k_{\rm x}$ direction
(Figs.~\ref{fig:fig2}(c) and Fig.~\ref{fig:fig2}(d)), but also along
$k_{\rm y}$ direction (Fig.~\ref{fig:fig2}(e)). At $k_{\rm
y}$=1.65~\AA$^{-1}$ and $k_{\rm y}$=1.7~\AA$^{-1}$
(Figs.~\ref{fig:fig2}(c) and~\ref{fig:fig2}(d)), discontinuities of
the G band are observed at the crossing points with the Yb/G
($\beta_2^{\ast}$ and $\beta_1^{\ast}$) bands around $-$1.1~eV below
$E_{\rm F}$. At $k_{\rm x}$=0.0~\AA$^{-1}$ (Fig.~\ref{fig:fig2}(e)),
weak spectral intensity of the G band is observed at the crossing
points with the Yb/G bands around $-$0.5~eV and $-$1.1~eV below
$E_{\rm F}$ denoted by $E_{\rm hyb.}$ with white dashed lines.

\section{Discussions}

%-------------------Figure 3: Renormalization ------------------------%
\begin{figure}[b]
\includegraphics[width=1\columnwidth]{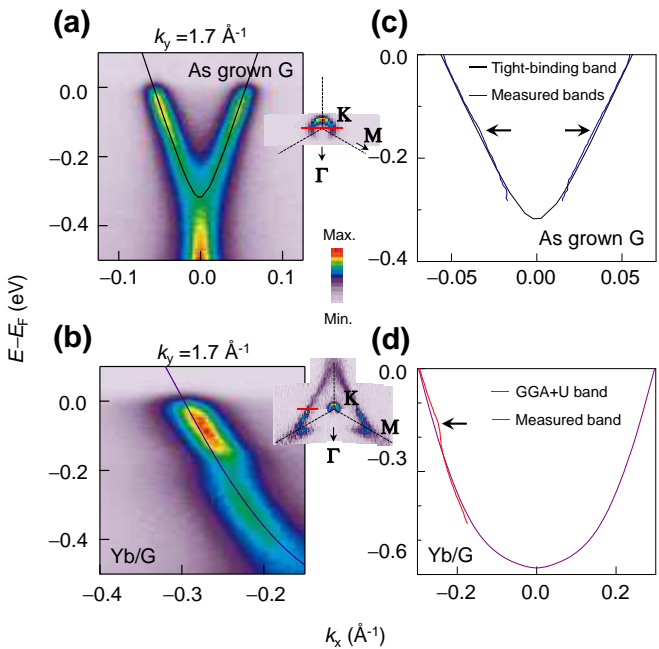}
\caption{\label{fig:fig3} (Color online) (a-b) Raw ARPES data for
the as grown G (panel (a)) and Yb/G ($\beta_1^{\ast}$: panel (b))
samples near $E_{\rm F}$ along the direction denoted by the red line
in the inset. (c-d) Comparison of measured and calculated bands of
as grown G (panel (c)) and Yb/G (panel (d)). The deviation at low
energy in the range of 0.2~eV from $E_{\rm F}$ is denoted by
arrows.}
\end{figure}
%-----------------------------------------------------------------

The calculated electronic band structure provides another important
information on the Yb/G system, i.\,e.\,, the $\pi$ bands of the
Yb/G crossing $E_{\rm F}$ ($\beta_1^{\ast}$) exhibits non-zero
contribution from Yb 6$s$ electrons in addition to the heavy carbon
$\pi$ character. In order to understand the impact of this
hybridization on the electronic properties, we investigate energy
spectra measured near $E_{\rm F}$ in comparison to calculated bands.
Figures~\ref{fig:fig3}(a) and~\ref{fig:fig3}(b) are raw ARPES data
of as grown graphene and Yb/G samples, respectively, along the
direction denoted by the red line in the inset of each panel. To
compare the measured and calculated bands quantitatively, we extract
energy-momentum dispersions using the standard method, i.\,e.\,,
Lorentzian fit to the momentum distribution curves (MDCs). The
measured band of as grown graphene is well described by the
tight-binding band, the black curve in Figs.~\ref{fig:fig3}(a)
and~\ref{fig:fig3}(c). On the other hand, the measured Yb/G band
($\beta_1^{\ast}$) shows a clear kinked structure around 0.16~eV
below $E_{\rm F}$ as denoted by an arrow in Fig.~\ref{fig:fig3}(d),
which is not expected in the GGA$+$U band, the purple curve in
Figs.~\ref{fig:fig3}(b) and~\ref{fig:fig3}(d). A similar structure,
although much weaker, is also observed in the G bands (arrows in
Fig.~\ref{fig:fig3}(c)). Such a kinked structure has been
extensively studied in the literature in the context of band
renormalization due to the interaction of electrons with
phonons~\cite{Verga,Reinert,Hengsberger,Lanzara,Gweon}.

\begin{figure}[b]
\includegraphics[width=1\columnwidth]{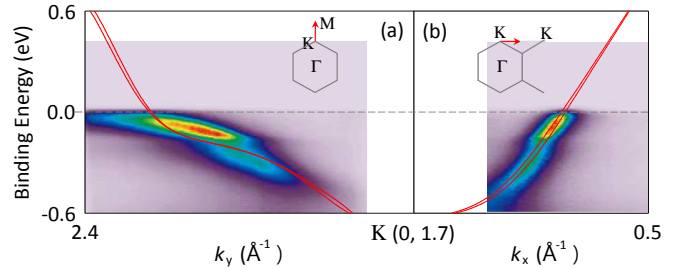}\\
\caption{\label{fig:fig4} (Color online) Measured and calculated
Yb/G band along the KM direction (panel (a)) and perpendicular to
the $\Gamma$K direction (panel (b)). The red curves are GGA$+$U
bands.}
\end{figure}

Before proceeding to a direct comparison between the effect of such
renormalization on G and Yb/G, and the consequent extraction of the
electron-phonon coupling constant, it is imperative to establish
whether these low energy kinked structures are real manifestation of
many body physics or just reflect the bare band structure of this
doped sample. Figure~\ref{fig:fig4} shows a comparison of the near
$E_{\rm F}$ band structure for Yb/G along the two directions (KM:
panel (a) and KK: panel (b)) with the GGA$+$U bands. Along the KM
direction (Fig.~\ref{fig:fig4}(a)), the measured band structure
clearly shows a kinked structure around 0.16~eV below $E_{\rm F}$.
However, GGA$+$U calculations (red curves) also show curved band
structure near the kink energy, which is not observed from the
electronic band structure of clean graphene, but induced due to an
hybridzation between the adsorbate electrons and the graphene $\pi$
bands. When the strength of electron-phonon coupling is determined
by the slope of the dispersion below and above the kink energy, this
curved band structure results in finite strength, despite the theory
does not include the electron-phonon coupling. This is the so-called
band structure effect~\cite{Calandra_BandStructureEffect}. In
addition, nearness to the van Hove singularity is supposed to spread
the measured spectral intensity away from the calculated Fermi
momentum, which is beyond the capability of our first principle
calculations. This spread out intensity results in the decrease of
the slope near $E_{\rm F}$ and hence the apparent enhancement of
electron-phonon coupling~\cite{Calandra_BandStructureEffect}.
Similar band structure effects have been extensively discussed in
the literature for Ca/G and
K/G~\cite{Jessica,Calandra_BandStructureEffect,Park_vanHove}. In
contrast, perpendicular to the  $\Gamma$K direction
(Fig.~\ref{fig:fig4}(b)), these non-trivial effects are not observed
allowing us to extract information on the electron-phonon coupling.

%------------------Figure 4: Band structure effect ------------------------%
%-----------------------------------------------------------------

%------------------Figure 5: Self-energy ------------------------%
\begin{figure}[t]
\includegraphics[width=1\columnwidth]{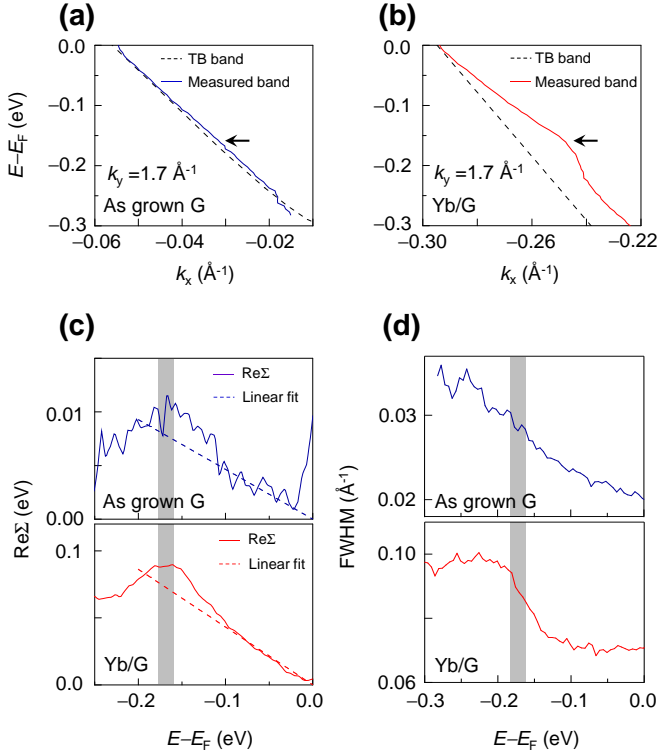}
\caption{\label{fig:fig5} %{\bf Electron self-energy enhanced by the presence of Yb}
(Color online) (a-b) Zoomed-in view of the electronic band structure
with a kink denoted by the arrow. The dashed line is a tight-binding
(TB) band fitted to each case. (c) Re$\Sigma$ for as grown G (upper
panel) and Yb/G (lower panel). The dashed line is a linear fit to
Re$\Sigma$ for $-$0.10~eV$\leq\,E-E_{\rm F}\leq$$-$0.03~eV. Shaded
area 0.16$\sim$0.18~eV below $E_{\rm F}$ is a peak position of the
Re$\Sigma$ spectrum. (d) MDC width as a function of $E-E_{\rm F}$
for as grown G (upper panel) and Yb/G (lower panel).}
\end{figure}
%-----------------------------------------------------------------

We now focus on the near $E_{\rm F}$ dispersion of as grown graphene
and Yb/G perpendicular to the  $\Gamma$K direction in
Figs.~\ref{fig:fig5}(a) and~\ref{fig:fig5}(b), respectively. It is
clear that, while the strength of the kink varies considerably, the
characteristic energy of the kink, 0.16~eV below $E_{\rm F}$, does
not change much. This implies a stronger coupling of electrons to
the optical phonon of graphene at the K point (A$_{1g}$ mode with an
energy $\hbar\omega_{\rm ph}$$\approx$0.16~eV) rather than the one
at the $\Gamma$ point (E$_{2g}$ mode with an energy
$\hbar\omega_{\rm ph}$$\approx$0.19~eV), in agreement with previous
reports for as grown graphene~\cite{Zhou08} and as expected in the
case of enhanced electronic correlations~\cite{Basko}. Similar
conclusion can be drawn from the real part of the electron
self-energy (Re$\Sigma$), i.\,e.\,, the difference between the
measured band and the tight-binding band, and from the imaginary
part of electron self-energy (Im$\Sigma$) which is proportional to
the full width at half maximum (FWHM) of
MDCs. %, where a peak and a step at the phonon energy are expected,
%respectively, in the simple case of the Debye model.
In Figs.~\ref{fig:fig5}(c) and~\ref{fig:fig5}(d), we report the
Re$\Sigma$ and FWHM spectra. Re$\Sigma$, in both cases, is dominated
by a strong peak at 0.16-0.18~eV (gray shaded area), while the FWHM
exhibits an enhanced quasiparticle scattering rate (or increased
width) around the same energy. The shape of Re$\Sigma$ and
Im$\Sigma$ for Yb/G is consistent with the theoretical prediction of
the electron-phonon coupling for highly electron-doped
graphene~\cite{Park2007}. The upturn of the Re$\Sigma$ spectra close
to $E_{\rm F}$ is a well-known resolution effect, which typically
results in the deflection of MDC peaks within a few tens meV near
$E_{\rm F}$ to lower momentum~\cite{Plumb,Valla2006}, and would
result in the apparent increase of Re$\Sigma$ close to $E_{\rm F}$.

\begin{figure}[t]
\includegraphics[width=1\columnwidth]{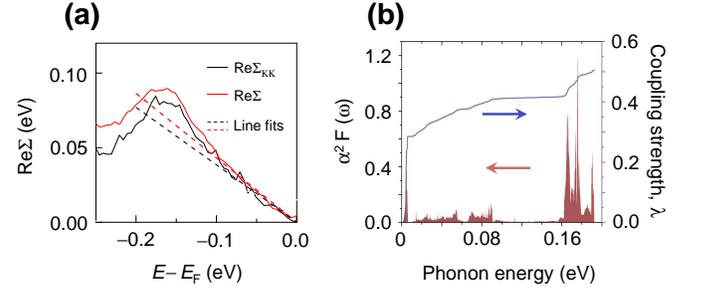}
\caption{\label{fig:fig6} (Color online) (a) Re$\Sigma$ (red line)
and Re$\Sigma_{\rm KK}$ (black line) for Yb/G. Dashed lines are
linear fits to Re$\Sigma$ and Re$\Sigma_{\rm KK}$ for
$-$0.10~eV$\leq\,E-E_{\rm F}\leq$$-$0.03~eV and
$-$0.1~eV$\leq\,E-E_{\rm F}\leq$0~eV, respectively. (b) Calculated
electron-phonon coupling spectrum $\alpha^2$F($\omega$) (brown
shaded area) and the evolution of $\lambda$ as a function of phonon
energy (navy curve) for Yb/G.}
\end{figure}

The real part of electron self-energy is a direct measurement of the
electron-phonon coupling constant, given by
$\lambda$=$\mid$$\partial{\rm Re}\Sigma(E)/\partial{E}$$\mid_{E_{\rm
F}}$. The dashed line in Fig.~\ref{fig:fig5}(c) is a linear fit to
Re$\Sigma$ for $-$0.10~eV\,$\leq E-E_{\rm F} \leq$\,$-$0.03~eV. We
obtain $\lambda$=0.046$\pm$0.002 for as grown graphene, which is
similar to the previously reported theoretical
($\lambda$=0.02)~\cite{Calandra_resolution} and experimental
($\lambda$=0.14)~\cite{Zhou08} values. The difference from the
latter might originate from the method to extract $\lambda$. For
Yb/G, we obtain $\lambda$=0.431$\pm$0.004, which exhibits strong
enhancement by an order of magnitude compared to the value for as
grown graphene. It is important to note that the GGA$+$U band in
Fig.~\ref{fig:fig3}(d) does not show the decreasing slope of the
dispersion near $E_{\rm F}$, so the band structure effect is safely
excluded as the origin of the enhanced
$\lambda$~\cite{Calandra_BandStructureEffect}. The self-consistency
of the self-energy analysis is obtained via Kramers-Kronig
transformation of Im$\Sigma$~\cite{Kordyuk} as shown in
Fig.~\ref{fig:fig6}(a). The strength of the electron-phonon coupling
is obtained by linear fits to Re$\Sigma$ and Re$\Sigma_{\rm KK}$
(brown dashed lines) for $-$0.10~eV$\leq\,E-E_{\rm F}\leq$$-$0.03~eV
and $-$0.1~eV$\leq\,E-E_{\rm F}\leq$0~eV, resulting in
$\lambda$=0.431$\pm$0.004 and $\lambda_{KK}$=0.385$\pm$0.011,
respectively.

%------------------Figure 6: KK transform and el-ph calc. ------------------------%
%-----------------------------------------------------------------

The calculated electron-phonon coupling spectrum and electron-phonon
coupling constant for Yb/G (shown in Fig.~\ref{fig:fig6}(b)) are
obtained from the density-functional perturbation theory using the
program \textsc{Quantum ESPRESSO}~\cite{Baroni}. The electronic
orbitals were expanded in a plane-wave basis set with a kinetic
energy cutoff of 75~Ry.  The Brillouin zone integrations in the
electronic and phonon calculations were performed using
Monkhorst-Pack~\cite{Monkhorst} meshes. We refer to meshes of
$k$-points for electronic states and meshes of $q$-points for
phonons.  The electron-phonon coupling matrix elements were computed
in the first Brillouin zone on a 18$\times$18$\times$1 $q$-mesh
using individual electron-phonon coupling matrices obtained with a
36$\times$36$\times$1 $k$-points mesh. The electron-phonon coupling
spectrum, $\alpha^2$F($\omega$), (brown shaded area in
Fig.~\ref{fig:fig6}(b)), can be divided into three regions: (i)
low-energy Yb-related modes up to 0.005~eV; (ii) carbon out-of-plane
modes up to 0.09~eV; and (iii) carbon in-plane modes at
0.16$\sim$0.18~eV and 0.19~eV. We find very strong electronic
coupling to the phonon mode at 0.16$\sim$0.18~eV in agreement with
our observation (see Fig.~\ref{fig:fig5}(c)). The coupling strength
can be directly determined from the spectra being
$\lambda$=2$\int$d$\omega \alpha^2$F($\omega$)/$\omega$ (navy curve
in Fig.~\ref{fig:fig6}(b)). Clearly the electron-phonon coupling
constant is drastically enhanced with respect to the as grown sample
over the entire range, from $\lambda$=0.02 for as grown
graphene~\cite{Calandra_resolution} to $\lambda$=0.51 for Yb/G,
consistent with the observed enhancement from $\lambda$=0.05 to
$\lambda$=0.43 (Fig.~\ref{fig:fig5}). The difference of the
experimental $\lambda$ from the theoretical value might be caused by
the lack of the exact unrenormalized band in extracting Re$\Sigma$,
which underestimates experimental $\lambda$~\cite{DavidNJP}.

The observed enhancement up to 0.43 (experimental) and 0.51
(theoretical) due to Yb is far greater than the theoretically and
experimentally estimated enhancement up to $\sim$0.09 by the change
of charge carrier density up to
$n$=1.7$\times$10$^{14}$~cm$^{-2}$~\cite{Calandra_resolution,DavidNJP},
as for the Yb/G sample. This indicates that charge doping alone
cannot explain the observed enhancement. Similar enhancement beyond
the capability of charge carrier density has been observed for
potassium-intercalated graphene on Ir substrate~\cite{Ref1,Ref2}
with $\lambda$=0.2$\sim$0.28. In the case of calcium-intercalate
graphene on Au/Ni(111)/W(110) substrate~\cite{Ref3}, the anisotropic
increase of $\lambda$ from 0.17 (along the $\Gamma$K direction) to
0.40 (along the KM direction) has been controversial as ascribed to
a change of the electron band structure and the van Hove singularity
due to the Ca intercalation, which result in apparent enhancement of
$\lambda$~\cite{Calandra_BandStructureEffect,Park_vanHove}.

The observed $\lambda$=0.43 in our work is the highest value ever
measured for graphene. It is interesting to note that, for bulk
graphite, the electron-phonon coupling in the Yb intercalated sample
(Yb-GIC) is estimated to be weaker than that in the Ca intercalated
sample (Ca-GIC), because of the slightly larger interlayer
separation which leads to a decrease of the interlayer-$\pi^{*}$
electron-phonon matrix element and thus smaller superconducting
phase transition temperature, $T_{\rm c}$ (6.5~K for Yb-GIC versus
11.5~K for Ca-GIC~\cite{Weller}). This trend is reversed in their
graphene counterparts, $\lambda$=0.43 for Yb/G (in this work) versus
$\lambda$=0.4 (or 0.17) for Ca/G~\cite{Ref3} suggesting that the
hybridization between graphene $\pi$ bands and the electrons from
adatoms governs the low energy excitations in monolayer graphene. %In GICs,
%the interlayer state is proposed to be essential to enhance the
%electronic coupling to the carbon phonon modes compared to that of
%clean graphite~\cite{Profeta}. On the other hand, in Yb/G, we do not
%observe such an interlayer state in the electronic band structure,
%hence the observed enhancement of electron-phonon coupling is even
%more surprising, which suggests the important role of ytterbium in
%determining the electronic properties of graphene.
%In fact, the clear difference of the Yb/G bands from the as grown
%graphene bands is the hybridization of the adatom states with the
%graphene $\pi$ bands constituting the $\beta_1^{\ast}$ band that
%determines low-energy excitations in this system.
The hybridization induces strong Coulomb interactions, as evidenced
by the preeminent role of the K point phonon compared to the
$\Gamma$ point phonon in the electron-phonon coupling~\cite{Basko}
as shown in Figs.~\ref{fig:fig5} and~\ref{fig:fig6}, and allows
phonons to be strongly coupled to electrons in graphene.

In line with the plausible phonon-mediated superconductivity in
Yb-GIC, the strong enhancement of electron-phonon coupling in Yb/G
suggests the exciting possibility that the introduction of Yb might
induce superconductivity~\cite{Rose,McMillan}. The $T_{\rm c}$ is
calculated using the Allen-Dynes equation~\cite{Allen},
\begin{equation}
  T_{\rm c}= {\Omega_{log} \over 1.2}
  \exp \left(- \frac{1.04(1+\lambda)}{\lambda - \mu^{*}(1+0.62 \lambda)} \right).
\end{equation}
The normalized weighting function of the Eliashberg
theory~\cite{McMillan} is
\begin{equation}
  g(\omega) = \frac{2}{\lambda \omega} \alpha^{2} F(\omega).
\end{equation}
The parameter $\lambda$ is a dimensionless measure of the strength
of $\alpha^{2} F$ with respect to frequency $\omega$:
\begin{equation}
  \lambda = 2 \int^{\omega}_{0} d \omega^{'}\,  \alpha^{2} F(\omega^{'}) / \omega^{'},
\label{eqn3}
\end{equation}
and the logarithmic average frequency, $\Omega_{log}$ in units of K,
is
\begin{equation}
  \Omega_{log} = \exp \left( \int^{\infty}_{0} g(\omega) \ln \omega \, d\omega  \right).
\end{equation}
The predicted $T_{\rm c}$ and $\Omega_{log}$ are estimated to be
1.71~K and 168.2~K, respectively. We use $\mu^{*}$=0.115 for proper
comparison with another theoretical work~\cite{Profeta} and it is
worth to note that the predicted $T_{\rm c}$ can range from 2.17~K
($\mu^{*}$=0.10) to 1.33~K ($\mu^{*}$=0.13).

%the interlayer state is proposed to be essential to enhance the
%electronic coupling to the carbon phonon modes compared to that of
%clean graphite~\cite{Profeta}. the important role of electron-phonon
%coupling in determining electronic properties has been proposed via
%the isotope effect, so that heavy element Yb induces lower
%superconducting phase transition temperature than lighter element Ca
%(6.5~K for Yb-GIC versus 11.5~K for Ca-GIC~\cite{Weller}). In other
%words, electron-phonon coupling of Yb-GIC is suggested to be weaker
%than that of Ca-GIC.  This trend is reversed in their graphene
%counterparts, e.\,g.\, $\lambda$=0.43 for Yb/G versus $\lambda$=0.4
%for Ca/G~\cite{Ref3}.  In GICs, the interlayer state is proposed to
%be essential to enhance the electronic coupling to the carbon phonon
%modes compared to that of clean graphite~\cite{Profeta}.

\section{Summary}
We have reported experimental evidence of strong enhancement of
electron-phonon coupling in graphene by as much as a factor of 10
upon the introduction of Yb (from 0.02$\leq$$\lambda$$\leq$0.05 to
0.43$\leq$$\lambda$$\leq$0.51). Such an enhancement goes beyond what
one would expect by charge doping. Our results reveal the important
role of the hybridization between electrons from Yb adatoms and the
graphene $\pi$ electrons, pointing to such hybridization as a
critical parameter in realizing correlated electron phases in
graphene.
%The comparison between measured and calculated band structures
%allows us to exclude the band structure effect as the origin of the
%enhanced electron-phonon coupling, and reveals that the
%hybridization between Yb and graphene electrons plays an important
%role in the enhancement beyond the simple charge doping case.

\acknowledgments The experimental part of this work was supported by
Berkeley Lab's program on sp2 bond materials, funded by the U.S.
Department of Energy, Office of Science, Office of Basic Energy
Sciences, Materials Sciences and Engineering Division, of the U.S.
Department of Energy (DOE) under Contract No.~DE-AC02-05CH11231.
%CK
%C.\ H.\ was supported by the Research Fund Program of Research
%Institute for Basic Sciences, Pusan National University, Korea,
%2013, Project No.~RIBS-PNU-2013-311.
%Jeff
Work at the Molecular Foundry was supported by the Office of
Science, Office of Basic Energy Sciences, of the U.S. Department of
Energy under Contract No.~DE-AC02-05CH11231.
%Cohen
The theoretical part of this work was supported by NSF Grant No.
DMR-IO-1006184 and the theory program at the Lawrence Berkeley
National Laboratory through the Office of Basic Energy Science, US
Department of Energy under Contract No.~DE-AC02-05CH11231 (K.\ T.\
C.\,, J.\ N.\,, and M.\ L.\ C.); and by the Energy Frontier Research
in Extreme Environments Center (EFree) under award number
DE-SG0001057. B.\ J.\ acknowledges financial support from the
European Research Council (ERC-2008-AdG-No.~228074).

%\section*{References}

\end{document}